\documentstyle[aps,prb,multicol,graphics]{revtex}

\begin{document}
\title{Electron-phonon interaction in the normal and superconducting states of MgB$%
_{2}$}
\author{Y. Kong, O.V. Dolgov, O. Jepsen, and O.K. Andersen}
\address{Max-Planck-Institut f\"{u}r Festk\"{o}rperforschung,\\
Stuttgart, Germany}
\date{\today}
\maketitle

\begin{abstract}
For the 40K-superconductor MgB$_{2}$ we have calculated the electronic and
phononic structures and the electron-phonon interaction throughout the
Brillouin zone {\it ab initio}. In contrast to the isoelectronic graphite,
MgB$_{2}$ has holes in the bonding $\sigma $-bands, which contribute 42 per
cent to the density of states: $N\left( 0\right) =0.355$\thinspace states/(MgB%
$_{2}\cdot $eV$\cdot $spin). The total interaction strength, $\lambda =0.87$
and $\lambda _{tr}=0.60,$ is dominated by the coupling of the $\sigma $%
-holes to the{\bf \ }bond-stretching optical phonons with wavenumbers in a
narrow range around 590 cm$^{-1}.$ Like the holes, these phonons are quasi
two-dimensional and have wave-vectors close to $\Gamma $A, where their
symmetry is E. The $\pi $-electrons contribute merely 0.25
to $\lambda $ and to $\lambda _{tr}.$ With Eliashberg theory we evaluate the
normal-state resistivity, the density of states in the superconductor, and
the B-isotope effect on $T_{c}$ and $\Delta _{0},$ and find excellent
agreement with experiments, when available. $T_{c}$=40\thinspace K is
reproduced with $\mu ^{\ast }=0.10$ and $2\Delta _{0}/k_{B}T_{c}=3.9.$ MgB$%
_{2}$ thus seems to be an intermediate-coupling
e-ph pairing $s$-wave superconductor.
\end{abstract}

\pacs{63.20.Kr, 74.20.-z, 74.25.Kc, 74.25.-q}

\begin{multicols}{2}

The recent discovery\cite{Aki} of superconductivity with $T_{c}$=39{\rm %
\thinspace }K in the graphite-like compound MgB$_{2}$ has caused hectic
activity. Density-functional (LDA) calculations\cite{kortus,Bela,Satta,AP}
show that, in contrast to intercalated graphite $\left( T_{c}\leq 5\,{\rm K}%
\right) $ and alkali-doped fullerides, A$_{3}$C$_{60}$ $\left( T_{c}<40\,%
{\rm K}\right) ,$ in MgB$_2$ there are holes at the top of the 
B-B bonding $\sigma $%
-bands, and that these couple rather strongly to optical B-B bond-stretching
modes\cite{AP} with wavenumbers\cite{kortus,Satta} around 600\thinspace cm$%
^{-1}$. These are the same type of modes as those believed to couple most
strongly to the $\pi $-electrons in graphite and C$_{60}$, where their
wavenumbers are 2.5 times larger, however\cite{Chan,Gunn}. Rough estimates
\cite{kortus,AP} of the electron-phonon coupling-strength for $s$-wave
pairing in MgB$_{2}$ yield: $\lambda \sim $1. Measurements of the B-isotope
effect\cite{budiso} on $T_{c}$, tunnelling\cite{sharoni}, transport \cite
{finnemore,jung,canfield,bud}, thermodynamic properties\cite
{finnemore,Kremer,Ott}, and the phonon density of states\cite{sato,Osborn}
confirm that MgB$_{2}$ is most likely an electron-phonon mediated $s$-wave
superconductor with intermediate\cite{sharoni,Kremer,Osborn} or strong\cite
{Ott} coupling.

In order to advance, detailed comparisons between accurate 
results of Eliashberg
theory and experiments are needed. Consider again the example of A$_{3}$C$%
_{60},$ also believed to be conventional $s$-wave superconductors\cite{Gunn}
with $T_{c}$'s described by the McMillan expression: 
\begin{equation}
T_{c}^{McM}=\frac{\omega _{\ln }}{1.2}\exp \left[ \frac{-1.04(1+\lambda )}{%
\lambda -(1+0.62\lambda )\mu ^{\ast }}\right] .  \label{McM}
\end{equation}
The LDA values for $\lambda $ are 0.4--0.6 and the value of the Coulomb
pseudopotential $\mu ^{\ast }$ to be used in (\ref{McM}) is presumably
considerably larger than the usual value, 0.1--0.2$,$ for $sp$%
-materials\cite{savr} due to the small width ($\sim $0.5 eV) of the 
$\pi $ $t_{1u}$-subband compared with the on-ball Coulomb repulsion\cite{Gunn}.
For MgB$_{2},$ where the $\pi $-band is 15 eV broad, one expects $\mu
^{\ast }$ to be 0.1--0.2 and the LDA\cite{VWN} plus 
generalized-gradient correction\cite{GGA} to give $\lambda $ with an
accuracy better than 0.1.
The result of such a $\lambda $-calculation will be presented here and
should allow us to reach conclusions about  
the superconductivity in MgB$_2$.

MgB$_{2}$ consists of graphite-like B$_{2}$-layers stacked on-top with Mg
in-between. The primitive translations are: ${\bf a}$=$\left( \sqrt{3}%
/2,1/2,0\right) a,\;{\bf b}$=$\left( 0,a,0\right) ,\;{\bf c}$=$\left(
0,0,c\right) ,$ with\cite{lipp} $a=3.083${\rm \AA }$=5.826\,a_{0}$ and $%
c/a=1.142.$ In reciprocal space, and in units of $2\pi /a,$ the primitive
translations are: {\bf A}=$\left( 2/\sqrt{3},0,0\right) ,\;${\bf B}=$\left(
-1/\sqrt{3},1,0\right) ,\;${\bf C}=$\left( 0,0,a/c\right) ,$ and the points
of high symmetry are: ${\bf \Gamma }$=$\left( 0,0,0\right) ,$ {\bf A}=$%
\left( 0,0,a/2c\right) ,$ {\bf M}=$\left( 1/\sqrt{3},0,0\right) ,$ {\bf L}=$%
\left( 1/\sqrt{3},0,a/2c\right) ,$ {\bf K}=$\left( 1/\sqrt{3},1/3,0\right) ,$
{\bf H}=$\left( 1/\sqrt{3},1/3,a/2c\right) .$ 
To reach a numerical accuracy exceeding  0.1 for $\lambda $ requires
careful sampling $throughout$ the Brillouin-zone for electrons as well as
for phonons due to the small size of the cylindrical $\sigma $-hole
sheets\cite{kortus}. We therefore used Savrasov's\cite{savr} linear-responce
full-potential LMTO density-functional method, proven to describe the
superconducting and transport properties of $e.g.$ Al and Pb with high
accuracy.
The Brillouin-zone
integrations were performed with the full-cell tetrahedron method\cite
{blochl} with the ${\bf k}$-points placed on the $\left( {\bf A},{\bf B},%
{\bf C}\right) /24$ sublattice. 
For the valence
bands, a triple-kappa $spd$ LMTO basis set was employed and the Mg 2$p$%
-semicore states were treated as valence states in a separate energy window.
The charge densities and potentials were represented by spherical harmonics
with $l\leq 8$ inside the non-overlapping MT spheres, and by plane waves
with energies $\leq $ 201\thinspace Ry in the interstitial region.

The resulting electronic structure is practically identical with that of
previous calculations\cite{kortus,Bela,Satta,AP}. Near and below the Fermi
level there are two B $p_{z}$ $\pi $-bands and three quasi-2D B-B bonding $%
\sigma $-bands. The $\sigma $ and $\pi $ bands do not hybridize when $k_{z}$%
=0 and $\pi /c$. The $\pi $-bands lie lower with respect to the $\sigma $%
-bands than in graphite and have more $k_{z}$-dispersion due to the
influence of Mg, the on-top stacking, and the smaller $c/a$-ratio$.$ This
causes the presence of $p_{\sigma l}$=0.056 light and $p_{\sigma h}$=0.117
heavy holes near the doubly-degenerate top along $\Gamma $A of the $\sigma $%
-bands. For the density of states at $\varepsilon _{F}\equiv 0,$ we find: $%
N\left( 0\right) =N_{\sigma l}\left( 0\right) +N_{\sigma h}\left( 0\right)
+N_{\pi }\left( 0\right) =0.048+0.102+0.205$ $=0.355$ states/(MgB$_{2}\cdot $%
eV$\cdot $spin).

\begin{figure}
\centerline{\resizebox{5.0in}{!}{\rotatebox{-90}{\includegraphics{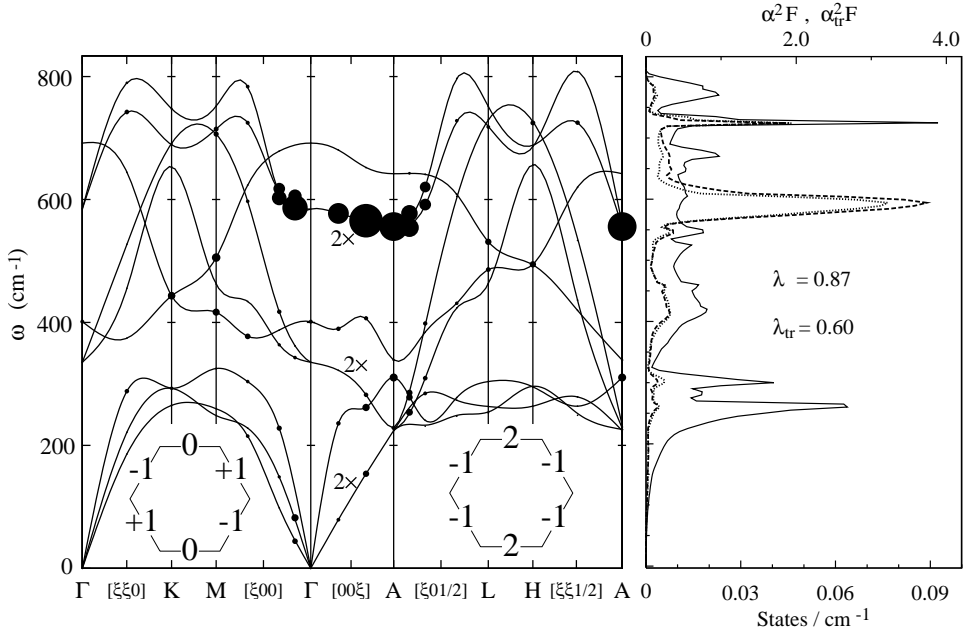}}}\\[2ex]}
\caption[]{\label{Phonon}
{\it Left:} Calculated phonon dispersion curves in MgB$%
_{2}$. The area of a circle is proportional to the mode-$\lambda .$ 
The insets at
the bottom show the two $\Gamma {\rm A}$ E eigenvectors (un-normalized), 
which apply to the holes
at the top of the $\sigma $-bands (bond-orbital coefficients)
as well as to the optical bond-stretching
phonons (relative change of bond lengths). 
{\it Right: }$F\left( \omega \right) $ (full curve and bottom
scale), $\alpha ^{2}\left( \omega \right) F\left( \omega \right) $ (broken)
and $\alpha _{tr}^{2}\left( \omega \right) F\left( \omega \right) $
(dotted). See text.}
\end{figure}

The $\sigma $ and $\pi $-bands may be understood and described with
reasonable accuracy near $\varepsilon _{F}$ using the orthogonal
tight-binding approximation with respectively the B $p_{z}$ orbitals and the
B-B two-center bond-orbitals formed from the B $sp^{2}$ hybrids: With two $%
p_{z}$ orbitals per cell and hopping between nearest neighbors only ($%
\varepsilon _{z}$=0.04\thinspace eV,\ $t_{z}^{\perp }$=0.92\thinspace eV,\
and $t_{z}$=1.60\thinspace eV$),$ the $\pi $-bands are\cite{kortus}: $%
\varepsilon _{\pi }\left( {\bf k}\right) =\varepsilon _{z}+2t_{z}^{\perp
}\cos ck_{z}\pm t_{z}\sqrt{1+4\cos \left( ak_{y}/2\right) \left[ \cos \left(
ak_{y}/2\right) +\cos \left( ak_{x}\sqrt{3}/2\right) \right] }.$ The bonding 
$\sigma $-band Hamiltonian is: 
\begin{eqnarray*}
&&H_{\sigma }\left( {\bf k}\right) =t_{sp^{2}}-2t_{b}^{\perp }\cos ck_{z} \\
&&-2t_{b}\left\{ 
\begin{array}{ccc}
0 & \cos \gamma +r\cos \left( \alpha +\beta \right)  & \cos \alpha +r\cos
\left( \beta +\gamma \right)  \\ 
c.c. & 0 & \cos \beta +r\cos \left( \alpha -\gamma \right)  \\ 
c.c. & c.c. & 0
\end{array}
\right\} ,
\end{eqnarray*}
in the representation of the three bond-orbitals per cell, with $t_{sp^{2}}$
being the energy of the two-center bond, and the integrals for hopping
between nearest and 2nd-nearest bond orbitals in the same layer being
respectively $t_{b}$=5.69\thinspace eV and $t_{b}^{\prime }\equiv rt_{b}$%
=0.91\thinspace eV, and with $t_{b}^{\perp }$=0.094\thinspace eV being an
order of magnitude smaller than $t_{z}^{\perp }$. Moreover, $\alpha \equiv 
\frac{1}{2}{\bf k\cdot a,\;}\beta \equiv \frac{1}{2}{\bf k\cdot b,}$ and $%
\gamma \equiv \frac{1}{2}{\bf k\cdot }\left( {\bf b-a}\right) .$ Along $%
\Gamma $A, $\alpha $=$\beta $=$\gamma $=$0$ so that there is a
singly-degenerate band of symmetry A with dispersion $t_{sp^{2}}-2t_{b}^{%
\perp }\cos ck_{z}-4\left( t_{b}+t_{b}^{\prime }\right) $ and a
doubly-degenerate band of symmetry E with dispersion $t_{sp^{2}}-2t_{b}^{%
\perp }\cos ck_{z}+2\left( t_{b}+t_{b}^{\prime }\right) .$ The E-band is
slightly above the Fermi level and its eigenvectors are given in the two
inserts at the bottom of Fig. \ref{Phonon}. The Fermi-surface sheets are
warped cylinders\cite{kortus} which may be described by expanding the two
upper bands of $H_{\sigma }\left( {\bf k}\right) $ to lowest order in $%
k_{x}^{2}+k_{y}^{2}\equiv k_{\parallel }^{2}.$ This yields: 
\begin{equation}
\varepsilon _{\sigma n}\left( {\bf k}\right) =\varepsilon _{0}-2t_{b}^{\perp
}\cos ck_{z}-k_{\parallel }^{2}/m_{\sigma n}\,,  \label{esigma}
\end{equation}
where $\varepsilon _{0}\equiv t_{sp^{2}}+2\left( t_{b}+t_{b}^{\prime
}\right) =0.58$\thinspace eV is the average energy along $\Gamma $A and the
units of $k_{\parallel }^{2}/m_{\sigma n}$ and $k_{\parallel }$ are
respectively Ry and $a_{0}^{-1}.$ The light and heavy-hole masses are
respectively $m_{\sigma l}=4/\left( t_{b}a^{2}\right) =0.28$ and $m_{\sigma
h}=4/\left( 3t_{b}^{\prime }a^{2}\right) =0.59$ relatively to that of a free
electron. For energies so closely below $\varepsilon \left( \Gamma \right) $
that Eq. (\ref{esigma}) holds, the number-of-states function is $\int
k_{\parallel \,n}^{2}\left( \varepsilon ,k_{z}\right)
dk_{z}/k_{BZ}^{2}=k_{\parallel \,n}^{2}\left( \varepsilon ,\pi /2c\right)
/k_{BZ}^{2}=\left( \varepsilon _{0}-\varepsilon \right) m_{\sigma
n}/k_{BZ}^{2},$ and its energy derivative is therefore constant: $N_{\sigma
n}\left( \varepsilon \right) =m_{\sigma n}/k_{BZ}^{2}.$ Here, $\pi k_{BZ}^{2}
$=$\left( 2\pi a_{0}/a\right) ^{2}2/\sqrt{3}$ is the area, and $k_{BZ}$ the
average radius of the Brillouin zone. Note that the $\sigma $-sheets are
quite narrow: $k_{F\parallel \,l}\left( \pi /2c\right) /k_{BZ}$=$\sqrt{%
0.056/2}$=$0.17$ and $k_{F\parallel \,h}\left( \pi /2c\right) /k_{BZ}$=$%
\sqrt{0.117/2}$=$0.24.$ 

The dynamical matrix was calculated for {\bf q-}points on the $\left( {\bf A}%
,{\bf B},{\bf C}\right) /6$ sublattice using the B-mass 10.811 which
corresponds to the natural mix of isotopes. The phonon dispersions $\omega
_{m}\left( {\bf q}\right) $ and density of states $F\left( \omega \right) $
are shown in Fig.\ref{Phonon}. The agreement between our $F\left( \omega
\right) $ and those obtained from inelastic neutron scattering\cite
{sato,Osborn} is excellent; our peaks at 260 and 730\thinspace cm$^{-1}$ (32
and 90\thinspace meV) are seen in the experiments at 32 and 88\thinspace
meV. For the frequencies of the optical $\Gamma $-modes we get:
335\thinspace cm$^{-1}$ (E$_{1u}$), 401\thinspace cm$^{-1}$ (A$_{2u}$),
585\thinspace cm$^{-1}$ (E$_{2g}$), and 692\thinspace cm$^{-1}$ (B$_{1g}$).
These values are in good agreement with those of previous calculations,
except for the all-important E$_{2g}$ modes where the LAPW\cite{kortus} and
pseudo-potential\cite{Satta} calculations give values which are respectively 
$\sim $100\thinspace cm$^{-1}$ smaller and larger than ours. These doubly
degenerate modes are the optical B-B bond-stretching modes (obs). Close to $%
\Gamma $A, they have exactly the same symmetry and similar dispersions as
the light and heavy $\sigma $-holes, although with the opposite signs. The E
eigenfunctions shown at the bottom of Fig.\ref{Phonon} now refer to
displacement patterns, {\it e.g.,} $\left\{ -1,0,1\right\} $ has one bond
shortened, another bond stretched by the same amount, and the third bond
unchanged. These E displacement patterns will obviously modulate the
electronic bond energy, $t_{sp^{2}},$ in such a way that the light holes
couple to one, and the heavy holes to the other mode, with the same diagonal 
$\left( g_{\sigma ,obs}\right) $ and no off-diagonal matrix element. An and
Pickett\cite{AP} judged that this electron-phonon (e-ph) matrix element will
be the dominating one. This, we shall confirm, although we dispute their estimate of $%
\lambda $.

We now turn to the superconducting properties. Since the Fermi surface has
two $\sigma $ and two $\pi $-sheets, one might expect anisotropic pairing
with different gaps on different sheets. The experimental data\cite
{jung,canfield}, however, demonstrate that a change of residual resistivity
corresponding to a change of impurity scattering by two orders of magnitude
hardly affects $T_{c},$ although $T_{c}$ for anisotropic pairing is very
sensitive to impurity scattering. 
The relative linewidth
of the $m{\bf q}$-phonon due to e-ph coupling is\cite{AM}: 
\begin{eqnarray*}
\frac{\gamma _{m}\left( {\bf q}\right) }{\omega _{m}\left( {\bf q}\right) }
&=&2\pi \sum_{nn^{\prime }{\bf k}}\delta \left[ \varepsilon _{n}\left( {\bf k%
}\right) \right] \,\delta \left[ \varepsilon _{n^{\prime }}\left( {\bf k+q}%
\right) \right] \left| g_{n{\bf k,}n^{\prime }\,{\bf k+q},m}\right| ^{2} \\
&\equiv &\pi N\left( 0\right) \omega _{m}\left( {\bf q}\right) \lambda
_{m}\left( {\bf q}\right) ,
\end{eqnarray*}
where the factor 2 is from spin degeneracy and $\sum_{{\bf k}}$ is the
average over the Brillouin zone, so that $N\left( 0\right) $=$\sum_{n{\bf k}%
}\delta \left[ \varepsilon _{n}\left( {\bf k}\right) \right] .$ We have
(safely) assumed that $\omega _{m}\left( {\bf q}\right) \ll {\bf q\cdot v}%
_{n}\left( {\bf k}\right) ,$ where ${\bf v}_{n}\left( {\bf k}\right) \equiv 
{\bf \bigtriangledown }_{{\bf k}}\varepsilon _{n}\left( {\bf k}\right) $ is
the electron velocity. The e-ph matrix element is: $g_{n{\bf k,}n^{\prime }\,%
{\bf k+q},m}=\left\langle n{\bf k}\left| \delta V\right| n^{\prime }{\bf k}+%
{\bf q}\right\rangle /\delta Q_{m{\bf q}},\,$where the displacement in the $i
$-direction of the $j$th atom is related to the phonon eigenvector $e_{ij,m%
{\bf q}}$ and displacement $\delta Q_{m{\bf q}}$ by: $\delta R_{ij}=$ $%
e_{ij,m{\bf q}}\delta Q_{m{\bf q}}/\sqrt{2M_{j}\omega _{m{\bf q}}}.$ The
Eliashberg spectral function is:
\[
\alpha ^{2}\left( \omega \right) F\left( \omega \right) \equiv \frac{1}{2\pi
N\left( 0\right) }\sum_{m{\bf q}}\frac{\gamma _{m}\left( {\bf q}\right) }{%
\omega _{m}\left( {\bf q}\right) }\delta \left[ \omega -\omega _{m}\left( 
{\bf q}\right) \right] ,
\]
and the strength of the e-ph interaction is finally: $\lambda \equiv
2\int_{0}^{\infty }\omega ^{-1}\alpha ^{2}\left( \omega \right) F\left(
\omega \right) d\omega =\sum_{m{\bf q}}\lambda _{m}\left( {\bf q}\right) .$

\end{multicols}
 
\begin{multicols}{2}            

The dominance of the $\sigma $-$\sigma $ coupling via the optical
bond-stretching mode is clearly seen in Fig. \ref{Phonon}, where the area of
a black circle is proportional to $\lambda _{m}\left( {\bf q}\right) .$
Along $\Gamma {\rm A},$ 
except when ${\bf q\cdot v}<\omega ,$ only the small $k_{z}
$-dispersion $\left( t_{b}^{\perp }\right) $ makes $\lambda _{m}\left( {\bf q%
}\right) $ not diverge so that the numerical values are inaccurate due to 
the relative coarseness our ${\bf k}$%
-mesh. The nearly cylindrical $%
\sigma $-sheets, whose diameters are of about the same size as the smallest,
non-zero $q_{\parallel }$ on the affordable $\left( {\bf A},{\bf B},{\bf C}%
\right) /6$-mesh, require even more care in the numerical ${\bf q}$%
-integration: In case of a single cylindrical sheet with $p$ holes, $\lambda
\left( q_{\parallel }\right) $ has the well-known $%
\mathop{\rm Im}%
\chi \left( q_{\parallel }{\bf ,}\omega \rightarrow 0\right) $-form:$%
\;\lambda \left( q_{\parallel }\right) /\lambda =\left( 2\pi px\sqrt{1-x^{2}}%
\right) ^{-1}\theta \left( 1-x\right) $ with $x$=$q_{\parallel
}/2k_{F\parallel }.$ This function vanishes when $q_{\parallel }$%
\mbox{$>$}%
$2k_{F\parallel },$ has a flat minimum of value $\left( \pi p\right) ^{-1}$
near $q_{\parallel }$=$\sqrt{2}k_{F\parallel },$ and has integrable
divergencies at $q_{\parallel }$=$2k_{F\parallel }$ and 0$.$ The proper
average of $\lambda \left( q_{\parallel }\right) $ is $\lambda .$ This
means, that $\lambda \left( {\bf q}\right) $ calculated on a coarse mesh
scatter violently for small $\left| {\bf q}\right| ,$ but that weighting
with $\lambda /\lambda \left( q_{\parallel }\right) $ gives the same,
correct result for all these points, provided that warping, as well as ${\bf %
k,k}^{\prime }$-dependence of $g$ and $\omega ,$ are neglected. In case of
two cylindrical sheets, and no coupling between them, $\lambda _{n}\left(
q_{\parallel }\right) /\lambda _{n}$ should be weighted by $m_{n}^{2}/\left(
m_{l}^{2}+m_{h}^{2}\right) .$ In our numerical evaluation of the e-ph
interaction with the linear-response code\cite{savr} we discarded the values
of $\lambda _{m}\left( {\bf q}\right) $ with ${\bf q}$ along $\Gamma $A, and
added those on the $\left( {\bf A}/12{\bf ,B/}12{\bf ,C}/6\right) $-mesh for
which $\sqrt{2}k_{F\parallel l}\lesssim q_{\parallel }\lesssim \sqrt{2}%
k_{F\parallel h}.$ The result was: $\lambda $=0.62+0.25, where 0.62 was the
contribution from ${\bf q}$'s so small that $\sigma $-$\sigma $ coupling
occurs, and 0.25 was the contribution from the remaining part of ${\bf q}$%
-space, which must involve a $\pi $-sheet. Had we included the inaccurate $%
\lambda _{m}\left( {\bf q}\right) $-values along the $\Gamma $A-line, the $%
\sigma $-$\sigma $\ result would have been 0.72 instead of 0.62. The result
was finally checked by using the approximate $\lambda \left( q_{\parallel
}\right) /\lambda $ correction for the point ${\bf q}$=${\bf A}/12.$ This
yielded 0.58 instead of 0.62. In conclusion: $\lambda =0.87\pm 0.05=\left(
0.62\pm 0.05\right) +0.25\equiv \lambda _{\sigma }+\lambda _{\pi }.$

The Eliashberg function shown on the right-hand side of Fig. \ref{Phonon} is
dominated by the large $\sigma $-$\sigma $ peak around $\omega _{obs}$=$590\,
$cm$^{-1}$=73\thinspace meV. The facts that the $\sigma $-sheets are narrow,
warped cylinders whose coupling is dominated by intra-sheet coupling via the
optical bond-stretching mode, and that the coupling between $\sigma $-
and $\pi $-sheets is
negligible, lead to the following approximation: 
\begin{eqnarray*}
&&\alpha ^{2}\left( \omega \right) F\left( \omega \right) \approx \alpha
_{\pi }^{2}\left( \omega \right) F\left( \omega \right) \left[ N_{\pi
}\left( 0\right) /N\left( 0\right) \right] + \\
&&\left| g_{\sigma ,obs}\right| ^{2}\delta \left( \omega -\omega
_{obs}\right) \left[ N_{\sigma l}^{2}\left( 0\right) +N_{\sigma h}^{2}\left(
0\right) \right] /N\left( 0\right) ,
\end{eqnarray*}
where $\alpha _{\pi }^{2}\left( \omega \right) F\left( \omega \right) $ is
the usual expression, but with $\pi $-electrons only. In An's and Picket's
estimate\cite{AP}, $\lambda _{\sigma }$=0.95, a factor $\left[ N_{\sigma
l}^{2}\left( 0\right) +N_{\sigma h}^{2}\left( 0\right) \right] /N_{\sigma
}\left( 0\right) N\left( 0\right) =0.24$ appears to be missing. The
rigid-atomic-sphere estimate $\lambda $=0.7 by Kortus {\it et al.} is closer
to our value 0.87.

Knowing $\alpha ^{2}\left( \omega \right) F\left( \omega \right) $ and a
value of the Coulomb pseudopotential $\mu ^{\ast }\left( \omega _{c}\right) $%
, we solve the Eliashberg equation on the real frequency axis\cite{Shu}, and
obtain $T_{c}$=40K if $\mu ^{\ast }\left( \omega _{c}\right) $=0.14$.$
Taking retardation effects into account, we find $\mu ^{\ast }\equiv $ $\mu
^{\ast }\left( \omega _{c}\right) /\left[ 1+\mu ^{\ast }\left( \omega
_{c}\right) \ln \left( \omega _{c}/\omega _{\ln }\right) \right] =0.10,$
where $\omega _{\ln }$=504\thinspace cm$^{-1}$=62\thinspace meV is obtained
from: 0=$\int_{0}^{\infty }\ln \left( \omega /\omega _{\ln }\right) \omega
^{-1}\alpha ^{2}\left( \omega \right) F\left( \omega \right) d\omega ,$ and
the cut-off frequency is taken as $\omega _{c}$=$10\max \omega $%
=8000\thinspace cm$^{-1}$. This value of $\mu ^{\ast }$ is at the lower end
of what is found for simple $sp$-metals\cite{savr}. The relation back to a
screened Coulomb interaction $U$ is: $\mu ^{\ast }=\mu /\left[ 1+\mu \ln
\left( \omega _{p}/\omega _{\ln }\right) \right] ,$ where $\mu $=$UN\left(
0\right) $ and $\omega _{p}\sim $7\thinspace eV is the plasma frequency
given below. We thus find: $\mu $=0.19 and $U$=1.1\thinspace eV, which are
normal values. Had we used the approximate McMillan expression (\ref{McM}),
the slightly higher value $\mu ^{\ast }$=0.14 would be needed to reproduce
the experimental $T_{c}.$

\begin{figure}
\centerline{\resizebox{2.5in}{!}{\rotatebox{-90}{\includegraphics{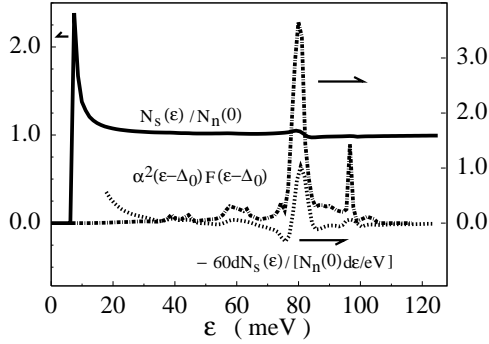}}}\\[2ex]}
\caption[]{\label{dos}
Normalized density of states (full) and the negative of its
energy-derivative (dotted) as obtained from the Eliashberg equation with $%
\mu ^{\ast }$=0.10 and $T$=3K.}
\end{figure}        
 
In Fig.\ref{dos} we show our Eliashberg calculation with $\mu ^{\ast }$=0.10
of the density of states, $N_{s}(\varepsilon )/N(0)=%
\mathop{\rm Re}%
\left[ \varepsilon /\sqrt{\varepsilon ^{2}-\Delta _{{\rm 3K}%
}^{2}(\varepsilon )}\right] ,$ in the superconductor. The BCS singularity is
at $\varepsilon $=$\Delta _{{\rm 3K}}\left( 0\right) $=6.8\thinspace meV,
which is in accord with the 4.9-6.9\thinspace meV found in tunneling
experiments\cite{sharoni}. This yields: $2\Delta _{0}/k_{B}T_{c}$=3.9 which
is slightly higher than the BCS value of 3.52. The distinct feature near
80\thinspace meV corresponds to the peak in $\alpha ^{2}(\omega )F(\omega )$
at 73\thinspace meV, shifted by the 6.8\thinspace meV gap. The latter
function is also shown in the figure together with the measurable quantity $%
-d^{2}I/dV^{2}\sim -dN_{s}(\varepsilon )/d\varepsilon .$

We have calculated the change in $T_{c}$ upon isotope substitution of $^{11}B
$ for $^{10}B$ and get: $\delta T_{c}$=--1.7\thinspace K, which corresponds
to the exponent $-\delta \ln T_{c}/\delta \ln M_{B}$=0.46. This agrees well
with the measured\cite{budiso} value: $\delta T_{c}$=--1\thinspace K. For
the change of the gap, which may be measured in tunnelling and optical
experiments, we calculate: $\delta \Delta _{0}$=--1.9\thinspace cm$^{-1},$
which corresponds to the exponent $-\delta \ln \Delta _{0}/\delta \ln M_{B}$%
=0.38.

\begin{figure}
\centerline{\resizebox{2.5in}{!}{\rotatebox{-90}{\includegraphics{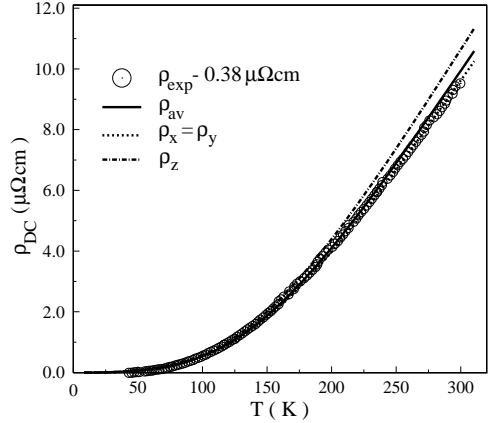}}}\\[2ex]}
\caption[]{\label{rho}
Calculated dc-resistivities in different directions compared
with the experiment in Ref. \onlinecite{canfield}.}
\end{figure}

Finally, we have considered transport properties in the normal state. Here,
solution of the kinetic equation leads to the transport e-ph spectral
function $\alpha _{tr,x}^{2}(\omega )F(\omega ),$ and similarly for $y$ and $%
z.$ These components are given by the previous expressions, but with the
additional factor $\left[ v_{nx}^{2}\left( {\bf k}\right) -v_{nx}\left( {\bf %
k}\right) v_{n^{\prime }x}\left( {\bf k+q}\right) \right] /\left\langle
v_{x}^{2}\right\rangle $ inserted. $\left\langle v_{x}^{2}\right\rangle
\equiv N\left( 0\right) ^{-1}\sum_{n{\bf k}}v_{nx}^{2}\left( {\bf k}\right)
\delta \left[ (\varepsilon _{n}\left( {\bf k}\right) \right] $. 
In Fig. \ref{Phonon} the directional
average, $\alpha _{tr}^{2}(\omega )F(\omega ),$ is seen to have the same
shape as $\alpha ^{2}(\omega )F(\omega ),$ except for the $\sigma $-$\sigma $
interaction via the optical bond-stretching modes, whose $\alpha
_{tr}^{2}(\omega )F(\omega )$ is smaller, presumably due to the near
two-dimensionality of the $\sigma $-bands. As a result, $\lambda _{tr}=0.60.$
\noindent For the plasma frequencies, $\omega _{p,x}^{2}=4\pi e^{2}N\left(
0\right) \left\langle v_{x}^{2}\right\rangle /\left[ {\bf abc}\right] ,$ we
find: $\omega _{p,x}$=$\omega _{p,y}$=7.02\thinspace eV and $\omega _{p,z}$%
=6.68\thinspace eV. Also the temperature dependence of the specific
dc-resistivity calculated with the standard Bloch-Gr\"{u}neisen expression, $%
\rho _{dc\,x}(T)=\left( \pi /\omega _{p,x}^{2}T\right) \int_{0}^{\infty
}\omega \sinh ^{-2}\left( \omega /2T\right) \alpha _{tr,x}^{2}(\omega
)F(\omega )d\omega ,$ is nearly isotropic and, as shown in Fig.\ref{rho}, is
in accord with recent measurements on dense wires\cite{canfield} over the
entire temperature range. The crossover from power-law to linear temperature
dependence is seen to occur near $\max \omega /5$=160\thinspace cm$^{-1}$%
=230\thinspace K, as expected\cite{allen86}.

In conclusion, we have presented an accurate {\it ab initio} calculation of
the e-ph interaction in MgB$_{2}$ and find $\lambda =0.87\pm 0.05.$
Eliashberg theory with $\mu ^{\ast }$=0.10 gives good agreement with
available experiments and several predictions. The unexpected high $T_{c}$
is due to the large $\lambda $-value caused by the presence of holes 
in the B-B bonding $\sigma $-band
and the relative softness of the optical bond-stretching modes. MgB$_{2}$
thus seems to be a simple and clear case of an intermediate-coupling e-ph
pairing $s$-wave superconductor.

Useful discussions with R. K. Kremer, I. I. Mazin, S. Savrasov, 
D. Savrasov, and S. V. Shulga are acknowledged.

\end{multicols}

\end{document}